\documentclass[aps,prl,twocolumn,preprintnumbers,amsmath,superscriptaddress,amssymb]{revtex4-1}
\usepackage{graphicx}
\usepackage{subfigure}
\usepackage{epsfig}
\usepackage{dcolumn}
\usepackage{bm}
\usepackage{ulem}
\usepackage{color}
\usepackage{float}
\usepackage{amsfonts}
\usepackage{amsmath}
\usepackage{bbm}
\usepackage{enumerate}
\usepackage[colorlinks,bookmarks=true,citecolor=blue,linkcolor=red,urlcolor=blue]{hyperref}

\def\abs#1{\left|#1\right|}

\def\be{\begin{equation}}       \def\ee{\end{equation}}
\def\bea{\begin{eqnarray}}      \def\eea{\end{eqnarray}}
\def\ba{\begin{array}}
	\def\ea{\end{array}}
\def\bnum{\begin{enumerate} }
	\def\enum{\end{enumerate}}

\def\=>{\Rightarrow}
\def\>{\rightarrow}

\def\eye2{Fathbb{I}}

\def\Fig#1{Fig.~\ref{#1}}

\renewcommand{\>}{\rangle}

\newcommand{\al}[1]{\begin{align}#1\end{align}}
\newcommand{\eq}[2]{
	\begin{equation}
		#1 \label{#2}
	\end{equation}
}

\newcommand{\re}[1]{\frac{1}{#1}}

\newcommand{\ipic}[4]{
\begin{figure}[h]\centering
	\includegraphics[#4]{#1}
	\caption{#2}  \label{#3}
\end{figure}
}

\begin{document}
\title{Universal properties of many-body localization transitions in quasiperiodic systems}
\author{Shi-Xin Zhang}
\affiliation{Institute for Advanced Study, Tsinghua University, Beijing 100084, China}
\author{Hong Yao}
\email{yaohong@tsinghua.edu.cn}
\affiliation{Institute for Advanced Study, Tsinghua University, Beijing 100084, China}
\affiliation{State Key Laboratory of Low Dimensional Quantum Physics, Tsinghua University, Beijing 100084, China}

\begin{abstract}
Precise nature of MBL transitions in both random and quasiperiodic (QP) systems remains elusive so far. In particular, whether MBL transitions in QP and random systems belong to the same universality class or two distinct ones has not been decisively resolved. Here we investigate MBL transitions in one-dimensional ($d\!=\!1$) QP systems as well as in random systems by state-of-the-art real-space renormalization group (RG) calculation. Our real-space RG shows that MBL transitions in 1D QP systems are characterized by the critical exponent $\nu\!\approx\!2.4$, which respects the Harris-Luck bound ($\nu\!>\!1/d$) for QP systems. Note that $\nu\!\approx\! 2.4$ for QP systems also satisfies the Harris-CCFS bound ($\nu\!>\!2/d$) for random systems, which implies that MBL transitions in 1D QP systems are {\it stable} against weak quenched disorder since randomness is Harris \textit{irrelevant} at the transition. We shall briefly discuss experimental means to measure $\nu$ of QP-induced MBL transitions.
\end{abstract}
\date{\today}
\maketitle

About one decade ago, Anderson localization in non-interacting systems \cite{Anderson1958, Fleishman1980, Johri2014} was generalized to many-body localization (MBL) in quantum many-body systems with interactions \cite{Basko2006, Oganesyan2007, Nandkishore2015, Altman2014, Vasseur2016, Znidaric2016,Abanin2017}. Isolated systems in MBL phases cannot thermalize where eigenstates thermalization hypothesis (ETH) \cite{Deutsch1991, Srednicki1994, Rigol2008} does not apply. In contrast to thermal systems, entanglement entropy of highly excited eigenstates of MBL systems obey the area law rather than volume law \cite{Bauer2013}, which renders various exotic properties such as extensive emergent integrable operators \cite{Serbyn2013a,Huse2014} and protected quantum order in excited states \cite{Huse2013,Potter2015a,Slagle2015}. More intriguingly, MBL transitions separating ergodic and MBL phases \cite{Pal2010a, Vosk2013, Pekker2014, Vasseur2015, Mondaini2015, Khemani2016a, Jian2017, Dai2018a, Stagraczynski2017} are so-called eigenstate phase transitions which can occur for each highly excited eigenstate with finite energy-density. For highly excited eigenstate across a MBL transition, entanglement entropy shifts from the volume law in thermal phases to the area law in MBL phases. This novel behavior cannot fit into the framework of conventional equilibrium phase transitions where entanglement entropy at finite temperature satisfies the volume law in both sides of transitions.
	
Currently there are two known mechanisms for MBL:  one by random disorder and the other by quasiperiodic (QP) potential. The former has been extensively studied for years, while the latter attracted increasing attentions very recently \cite{Iyer2013, Lee2017, Chandran2017, Nag2017, BarLev2017, Gray2017, Setiawan2017, Crowley2018,Znidaric2018} partly due to its accessibility in cold-atom experiments \cite{Schreiber2015a, Bordia2016, Luschen2017, Bordia2017}. Recently, an exact diagonalization (ED) study \cite{Khemani2017} indicated that MBL transitions in random systems and in QP systems belongs to two distinct university classes. Nonetheless, as ED studies are limited to models with relatively small size ($L$ up to about $20$), finite-size effect could be severe enough preventing to draw decisive conclusions. So far reliable study of universal properties of MBL transitions in QP models with much larger system size is still lacking. Therefore, it is highly desired to compute critical exponents by investigating systems with sufficiently large size to reliably address important questions such as: (i) whether QP MBL transitions is stable or not against weak randomness; (ii) whether MBL transitions in QP and random systems belongs to the same universality class.

\begin{figure}[t]
\includegraphics[width=0.8\columnwidth]{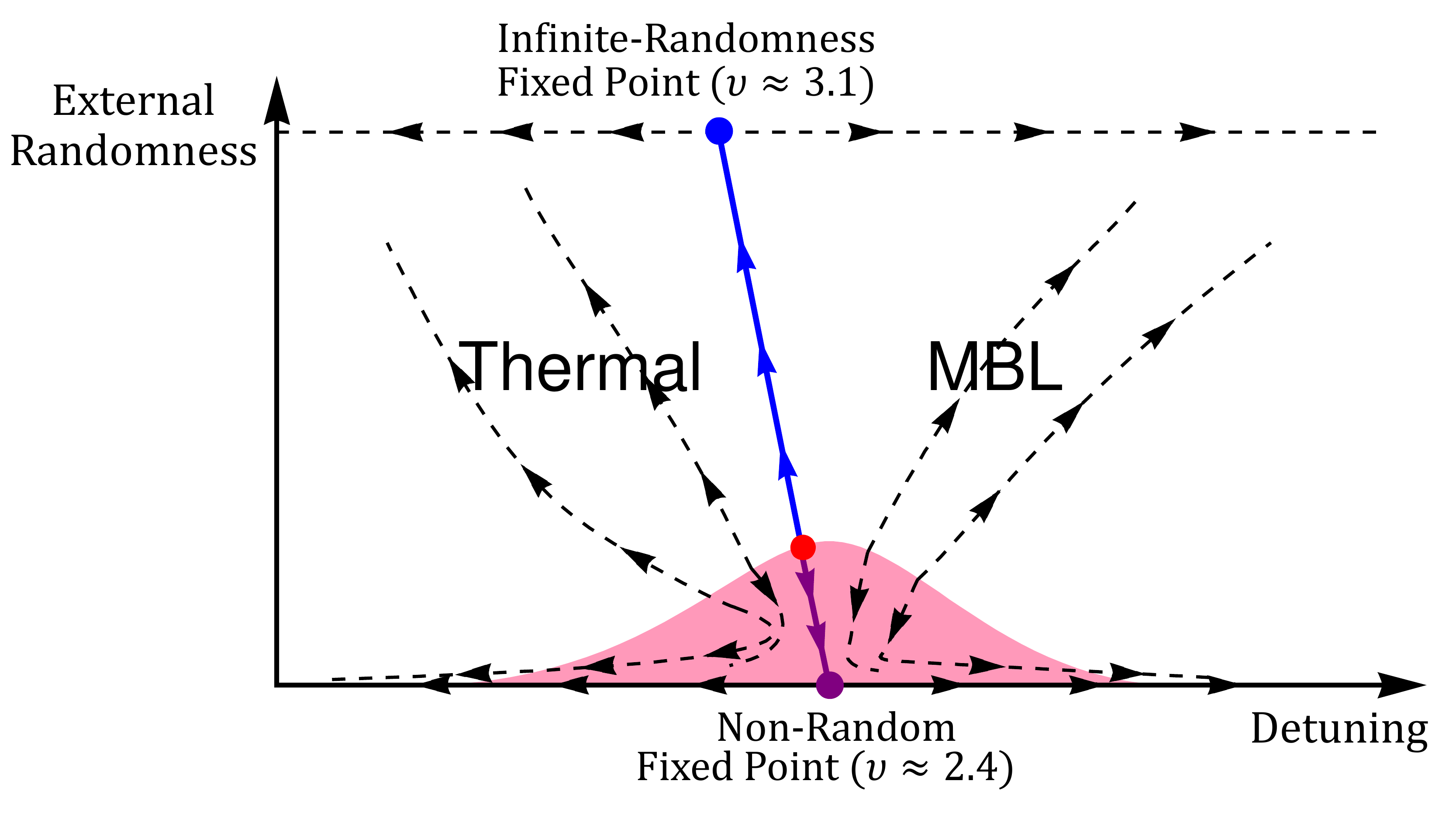}
\caption{The schematic RG flow of 1D interacting systems with both QP potential and quenched randomness. Our real-space RG analysis shows that MBL transitions induced by QP potential is Harris stable against weak randomness.}\label{RGflow}
\end{figure}
	
Novel real-space renormalization group (RSRG) approach was recently developed \cite{Vosk2015, Potter2015, Dumitrescu2017} to investigate critical behaviors of MBL transitions in random systems with much larger size compared to ED. It is probably the only numerical approach so far that obtained critical exponents satisfying the Harris-CCFS bound ($\nu\!>\!2/d$) \cite{harris1974, Chayes1986, Chayes1989, Luck1993a, Chandran2015} for randomness-driven MBL transitions. However, directly employing this approach to study MBL transitions in QP systems encounters serious problems. Here we propose a different RSRG approach by taking important microscopic details into account, which is more suitable to study MBL transitions in QP systems. By utilizing this improved approach, we can access 1D QP systems with size of order 1000, which is sufficiently large to investigate critical properties of MBL transitions. We systematically analyze different scenarios of MBL transitions using our RSRG approach, and come to conclusions that the critical exponent $\nu\!\approx\!2.4$ for QP-driven MBL transitions. This value of $\nu$ is in agreement with the Harris-CCFS bound ($\nu\!>\!2/d$), indicating that QP-induced MBL criticality can survive against weak randomness, as shown in the schematic RG flow in \Fig{RGflow}.

\textbf{Model:}	We consider the following one-dimensional spinless fermion model with interactions:
\bea\label{model}
H=-\sum_{ij}\left(t_{ij} c^\dagger _ic_j+H.c.\right)+\sum_i W_i n_i+V \sum_{\langle i j\rangle}n_in_j,~~
\eea
where $c^\dag_i$ creates a fermion at site $i$, $n_i=c^\dag_i c_i$ is the fermion density operator, $t_{ij}$ labels the hopping amplitude between sites $i$ and $j$, $V$ is the interaction between nearest-neighboring (NN) sites, and $W_i$ represents the onsite potential which can vary from site to site. Note that $W_i$ can be a random potential with uniform distribution $W_i\!\in\![0,W]$ for the random case or a cosine potential $W_i\!=\!W\cos(2\pi \alpha i+\phi)$ with quasi-periodicity $\alpha$ and phase $\phi$ (for simplicity, we set the irrational number $\alpha\!=\!\frac{\sqrt{5}-1}{2}$ as the golden ratio hereafter) for the QP case. Note that the QP potential has long-range correlation as $\langle W_i W_{i+j}\rangle=\frac12 W^2 \cos(2\pi\alpha j)$ which is in contrast to random potential without long-range correlation. In the limit of vanishing interaction $V\!=\!0$, the system with random potential is in the Anderson localization (AL) phase for any finite randomness $W\!>\!0$. For finite interaction $V$, the system with random potential would stay in the localized phase when the interaction $V$ is sufficiently weak but can go across the MBL phase transitions into an ergodic phase when $V$ exceeds a critical value. The nature of this MBL transition has been extensively studied \cite{Pal2010a, Vosk2013, Pekker2014, Vasseur2015, Khemani2016a}.

Here we shall focus on the case of QP potential. In the limit of vanishing interaction $V$, the Hamiltonian with QP potential $W_i$ is so-called Aubry-Andr\'e (AA) model \cite{Harper1955,Aubry1980}. The AA model has been extensively studied; it provides an example of single-particle localization in 1D with finite critical potential strength but without single-particle mobility edge. There have been various generalizations of the AA model \cite{Sarma1988,Biddle2009,Biddle2011a,Wang2013a,Ganeshan2015}. It turns out that generalized AA models usually possess single-particle mobility edge which indicates that the original AA model with only NN hopping is not generic in the family of QP models. Henceforth, we consider both NN hopping $t$ and next-nearest-neighborhood (NNN) hopping $t'$ such that this extended AA model could describe more realistic and generic systems \cite{Schreiber2015a, Bordia2016, Luschen2017, Bordia2017}. For this extended AA model with a sufficiently strong QP potential $W$, all single-particle states are localized. Then, increasing the interaction $V$ will presumably induce a MBL transition when $V$ exceeds a critical value $V_c$. We shall employ the real-space RG to study this putative MBL transition to address intriguing questions such as whether the MBL transition induced by QP potential is robust against weak random quenched disorder.

\textbf{Real-space RG approach:} In order to study the MBL transition in QP systems, we develop a RSRG approach by taking microscopic details into account, which goes beyond the previous RSRG approaches. We first briefly recapitulate the RSRG approach employed in studying the MBL transition in random systems \cite{Dumitrescu2017}. The RSRG method was constructed by mainly employing general nature of criticality, namely assumptions of scaling invariance near the MBL transition and the hierarchy of resonance clusters which can be implemented through iterations. To determine the structure of resonance clusters, one need to track the RG flow of two basic sets of simplified parameters between every two clusters: tunneling amplitude $\Gamma_{ij}$ (namely matrix element for transitions between different clusters) and typical energy mismatch $\Delta E_{ij}$.

For the initial condition or data input for  $\Gamma$ and $\Delta E$, the PVP RSRG assumes $\Delta E_{ij}=\abs{\mu_i-\mu_j}$ ($\mu_i$ is the onsite potential at $i$ and it takes box distribution for the case of random disorder) and $\Gamma_{ij}=V\exp{(-\abs{i-j}/{x_0})}$ ($x_0$ is the localization length and single-particle localization length is usually used as an approximation). Simply using chemical potential and sites positions as input data works well for studying MBL transitions in random systems, but encounters serious problems for studying MBL transitions in QP systems. In fact, if insisting to employ the original algorithm to study QP systems, one would obtain physical observables whose dependence with $W$ is not smooth. For reasons why the adiabatic approximation fails in QP systems, see Supplemental Materials.

To be capable of studying MBL transitions in QP systems, we improved the RSRG algorithm by making full use of the microscopic information of QP Hamiltonians. Specifically, we employ true spectrums of the single-particle Hamiltonian and localization centers of single-particle wave-functions as input data, rather than using the simplified ones. By doing this, we avoid the approximation caused by oversimplifications in the original approach. For technical details of the RSRG approach developed in the present work, see Supplemental Materials. To the best of our knowledge, it is the first time that critical exponents for MBL transitions in QP systems can be obtained from RG analysis of models with sufficiently large system size ($L$ order of 1000).

\begin{figure}[t]
\label{qp1}\includegraphics[width=0.8\columnwidth]{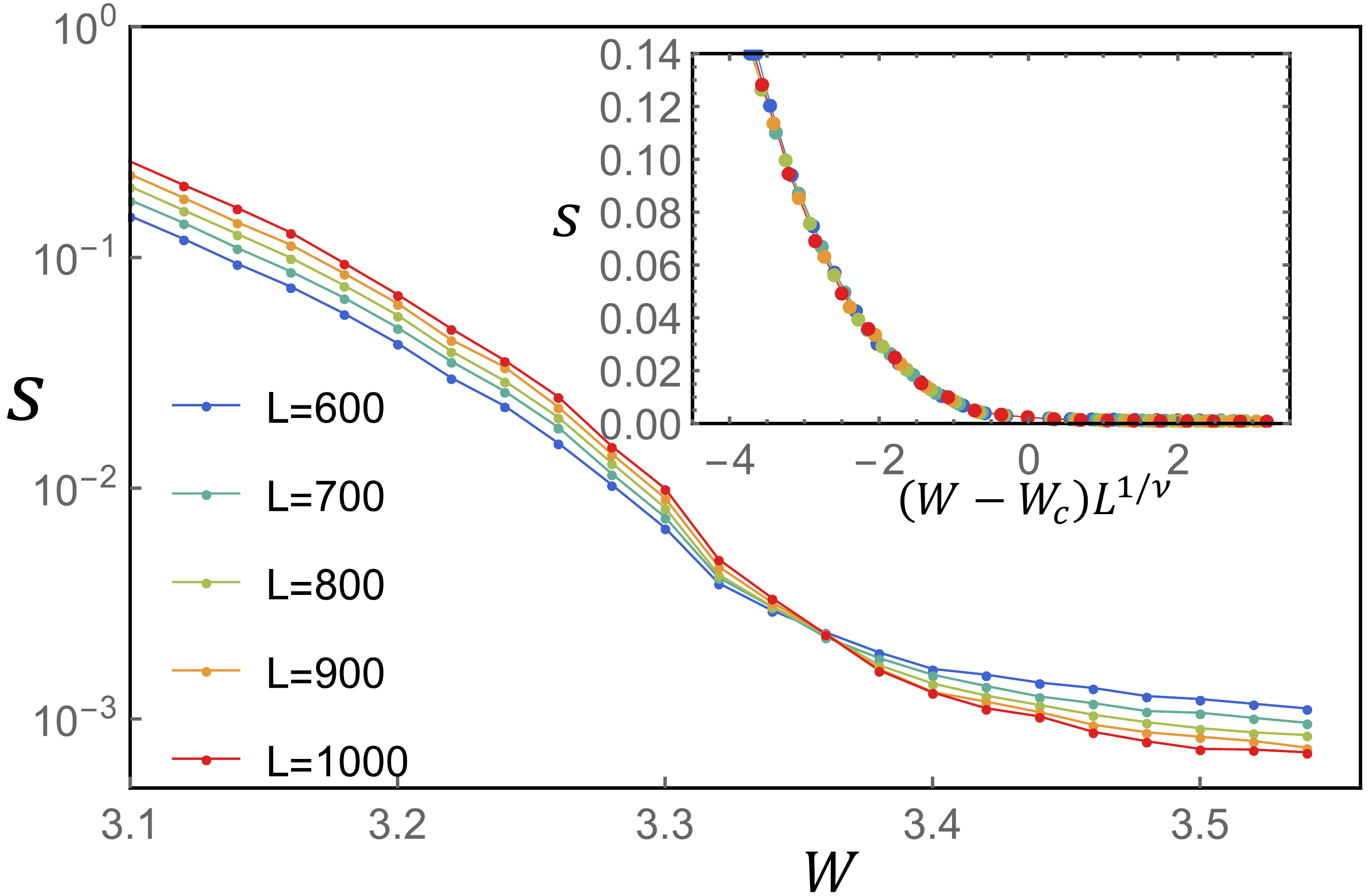}
\caption{Finite-size scaling analysis of entanglement entropy of 1D systems with quasiperiodic potentials $W_i\!=\!W\cos(2\pi \alpha i+\phi)$. The MBL transition is identified as the crossing point of entanglement entropy $s$ for different sizes $W_c\!=\!3.36\pm0.01$. The data collapse shown in the inset gives rise to $\nu\!=\!2.4\pm0.3$. Results are obtained by averaging over $10^5\!\sim\!10^6$ QP configurations (namely $10^5\!\sim\!10^6$ choices of $\phi$).} \label{fig-QP}
\end{figure}

\textbf{Results:} In performing real-space RG calculations of the interacting QP systems, we employ normalized entanglement entropy (EE) $s$ to analyze critical behaviors of the MBL transitions for the following two reasons. Firstly, entanglement entropy can be derived simply from the configurations and distributions of final resonance cluster structures; secondly, EE satisfies the scaling form as $s\!=\!f[(W-W_c)L^{1/\nu}]$ around MBL critical point $W_c$, where $f$ is some unspecified function, $L$ is the linear size of the system, and $\nu$ is the correlation (localization) length critical exponent.

Our RSRG results of normalized EE for various QP strength $W$ and various system size $L$ are shown in \Fig{fig-QP}, where we choose $t'\!=\!0.1t$ and $V=0.5t$. As mentioned before, the non-interacting extended AA model with finite NNN hopping $t'$ can have mobility edge in single-particle spectrum for $1.5\!\lesssim\!W\!\lesssim\!2.6$. Note that many-body mobility edges or exotic non-ergodic metal phase may occur between fully thermal and fully localized phases \cite{Modak2015, Li2015a, Li2016, Li2017a, Luschen2018, Naldesi2016} when weak interactions are added into non-interacting QP models with single-particle mobility edge (SPME). To understand the precise nature of such putative phases as well as the role SPME plays in such phases, more future works are desired.
Nonetheless, for sufficiently strong QP potential $W$ that can localize all single-particle states, a full MBL phase should naturally emerge when weak interactions are added. Indeed, as shown in \Fig{fig-QP}, the EE for all $L$ cross reasonably at the putative critical point $W\!=\!W_c\!\approx\! 3.36$ where all single-particle states are fully localized without single-particle mobility edge.

As mentioned above, the EE around the transition should satisfy the scaling form. Thus, we perform the scaling collapse, as shown in the inset of \Fig{fig-QP}, to extract the value of the critical exponent $\nu$ \cite{Houdayer2004}. The obtained result $\nu\!=\!2.4\pm0.3$ for the MBL transition in QP systems without quenched randomness is consistent with the Harris-Luck criterion ($\nu\!>\!1/d$) of QP systems \cite{Luck1993a}. To show the robustness of the obtained results, we also performed the same type of calculations for various settings of the model, including different interaction strength and different wave vectors. It is remarkable that critical exponents obtained for various settings are consistent with one another within error bar. This is not only a strong support to the numerical results we obtained, but also a clear demonstration of RSRG's effectiveness of capturing universal properties of MBL transitions. Moreover, this value of $\nu$ is also in agreement with the Harris-CCFS bound ($\nu\!>\!2/d$), implying that the MBL transition in QP systems is Harris stable against sufficiently weak randomness since the quenched randomness is Harris irrelevant at the QP MBL transition.

Note that the result of $\nu\!\approx\!2.4$ for QP MBL transitions in the present work is not in accord with the one in Ref. \cite{Khemani2017}, where $\nu\!\sim\!1$ was obtained through exact diagonalization (ED) calculations of models with relatively small system size ($L\le 18)$ and where the randomness was considered as Harris relevant due to the value of $\nu$ obtained by their ED calculations does not respect the Harris-CCFS bound ($\nu\!>\! 2/d$). The result of $\nu$ of the present paper derived from real-space RG calculations of models with system size up to $L\!=\!1000$ should have significantly less finite-size effect than ED calculations of models with system size ($L$ up to $20\!\sim\!30$). To directly verify that the quenched randomness is Harris-irrelevant at the QP-induced MBL transitions, we further carried out RSRG calculations on QP systems in the presence of a weak onsite randomness potential ($W'=0.1t$). With this weak quenched randomness, we obtained the critical value of QP potential $W_c=3.32\pm0.02$ as well as the critical exponent $\nu\!=\! 2.5\pm 0.3$. It is remarkable that the critical exponent of the QP-induced MBL transition in the presence of weak quenched randomness is consistent with the one without any quenched randomness, which confirms that the QP-MBL criticality is stable against weak randomness. Consequently, we reasonably conclude that weak randomness should be Harris irrelevant, rather than relevant, at the MBL transition of QP systems without quenched randomness, as shown in the schematic RG flow in \Fig{RGflow}.

\begin{figure}[t]
\label{r1}\includegraphics[width=0.8\columnwidth]{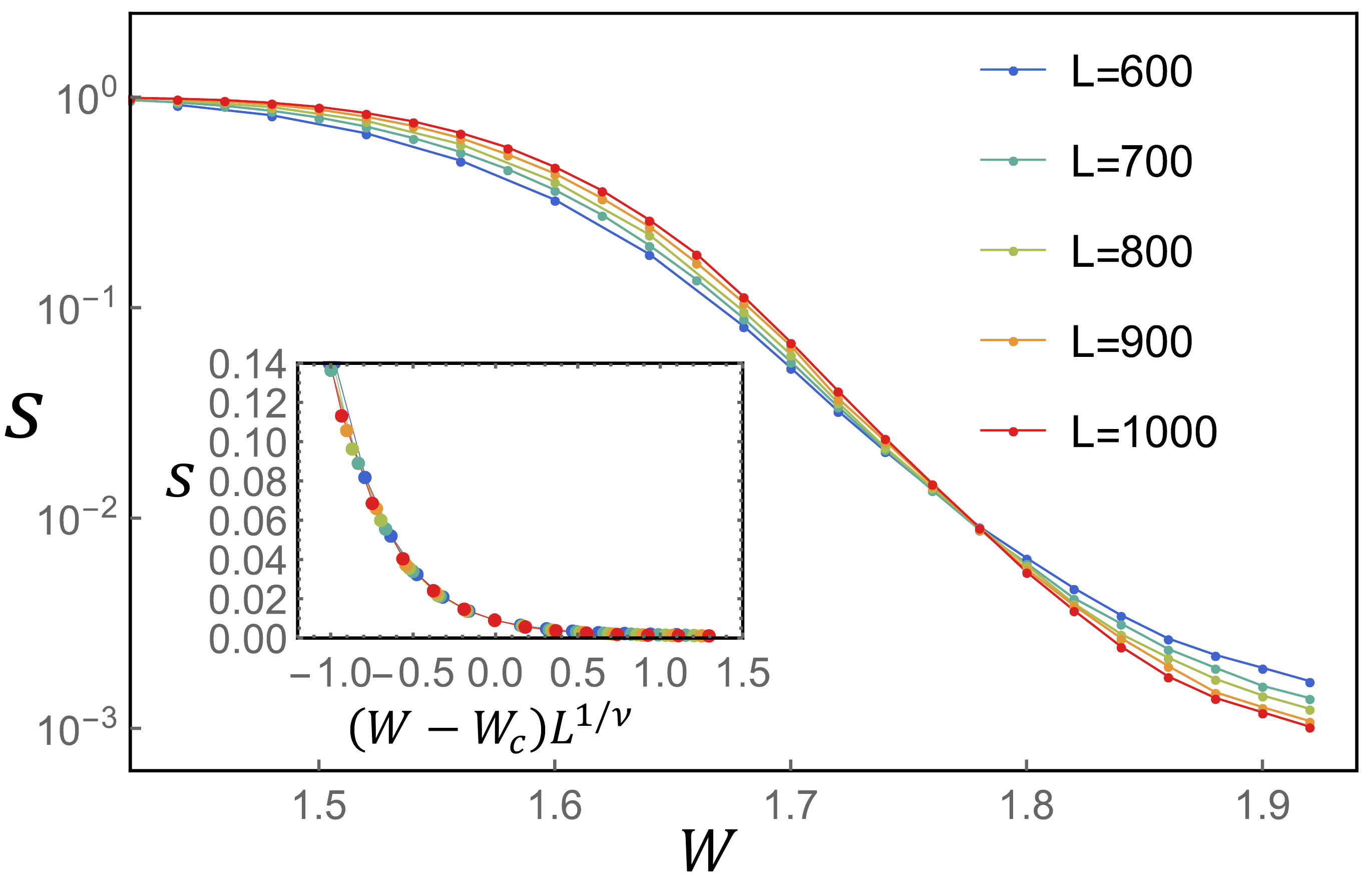}
\caption{The finite-size scaling for entanglement entropy $s$ of 1D system with random potentials $W_i\!\in\![0,W]$. The MBL transition occurs at $W_c\!=\!1.78\pm 0.01$. The data collapse shown in the inset gives rise to $\nu\!=\!3.1\pm 0.3$. Results are obtained by averaging over $10^5\!\sim\!10^6$ disorder configurations.} \label{fig-random}
\end{figure}

If MBL transitions induced by QP and by randomness belong to the same university class, they should feature the same $\nu$. Thus, we further performed RSRG calculations of models with quenched randomness; the results of normalized EE for different randomness $W$ and different system size $L$ are shown in \Fig{fig-random}. From the scaling collapse shown in the inset of \Fig{fig-random}, we obtained the critical exponent $\nu\!=\!3.1\pm 0.3$ for the MBL transitions induced by quenched randomness. This result is quite consistent with the one ($\nu\!=\!3.2\pm 0.3$) obtained by previous RSRG studies \cite{Vosk2015, Potter2015, Dumitrescu2017}, which indicates that our RSRG algorithm works well not only for QP systems but also for systems with quenched randomness. Since the result of $\nu\!\approx\!3.1$ for randomness-induced MBL transitions is significantly different from the result of $\nu\!\approx\!2.4$ of QP-induced MBL transitions, we believe that it is quite likely that they belong to two distinct universality classes. The results on critical behavior of both cases are summarized in TABLE \ref{resulttable}. Nonetheless, due to the difference between the values of $\nu$ for two cases are not that large after taking the errorbar into consideration, the RSRG results obtained in the present work could not fully rule out the possibility that MBL transitions driven by randomness and by QP potentials belong to the same universality class.

\textbf{Discussions and concluding remarks:} Note that the normalized EE curves of the QP systems shown in \Fig{fig-QP} are less smooth than the ones of random systems shown in \Fig{fig-random}. This may provide some insights to understand the difference between two seemingly distinct universality classes. We think that the less smooth behaviors of the QP case should be related to the multifractal properties of single-particle spectrum and wavefunctions of QP systems. In contrast to the random case where distribution of energy mismatch between neighboring localized single-particle eigenstates is continuous and smooth, the single-particle level mismatch between neighboring localized wavefunctions in QP systems possesses a self-similar feature with several gaps between sub-bands. Consequently, due to the fractal structure and gaps in the distribution of energy mismatch in QP systems, initial input data as well as final results of RSRG may show less smooth behaviours (even cusps) compared to the random case.

\begin{table}[t]
\centering
\caption{The results of critical exponents obtained from our improved algorithms of real-space RG for both randomness- and QP-induced MBL transitions.}
\label{resulttable}
\begin{tabular}{c|c|c}
 & \!QP-induced MBL\! & \!randomness-induced MBL\!
\\\hline
potential $W_i$   & $W\cos(2\pi\alpha i+\phi)$ & $[0, W]$
\\\hline
$\nu$ from RSRG & $2.4\pm 0.3$  & $3.1\pm 0.3$
\\\hline
\!lower-bound of $\nu\!$ & 1 & 2
\end{tabular}
\end{table}

We now discuss experimental ways to measure the critical exponent $\nu$ associated with the QP-induced MBL transitions, which could ultimately answer the question whether the QP MBL transition is stable or not against weak quenched randomness. One way of extracting the critical exponent $\nu$ is to experimentally measure transport properties in the Griffith region around MBL transitions. Specifically, the divergent behavior of dynamical exponent $z$ in the Griffith region around the MBL critical point $W_c$ is related with $\nu$: $z(W)\!\sim\!\re{(W-W_c)^\nu}$. $z\!>\!2$ for subdiffusive transport in Griffith region while $z\!=\!2$ for diffusive transport. Despite there are no randomness-induced rare regions in QP systems, Griffith region still appears around the QP-induced MBL transition \cite{Griffiths1969, Gopalakrishnan2016, Luschen2017} since rare regions can appear from preparing initial states. In Griffith region of 1D QP systems, subdiffusive behaviors should emerge. To probe subdiffusive transport experimentally, one can measure time dependence of density imbalance $I(t)$, which is defined as $I=\frac{N_e-N_o}{N_e+N_o}$ with $N_{e/o}$ the number of particles on even (odd) sites of 1D lattices. If the system is prepared with all atoms residing on even sites, the decaying behaviors of imbalance $I$ after a long time can be used as a sign to distinguish between ergodic and MBL phases. In the putative Griffith region, the imbalance $I$ should decay in a power law \cite{Potter2015, Luschen2017} $I(t)\sim t^{-1/z}$. Thus, by measuring time dependence of imbalance $I$, the Griffith region and the dynamic exponent $z(W)$ as a function of $W$ around MBL transition $W_c$ can be determined. From the behaviours of $z(W)$, one can extract $\nu$ which could ultimately help to determine the universality class of the MBL transitions in QP systems.

In conclusion, we improved the RSRG approach to make it more versatile to study models with qualitatively different disorders, especially suitable for studying MBL transitions in systems with QP potentials. Moreover, it paves one way for further investigations of MBL criticality in various types of models, such as the ones with longer-range hoppings, different dimensions, or different interactions. In the present paper, the critical exponent $\nu$ of MBL transitions in 1D QP systems obtained from our RSRG calculations satisfies the Harris-CCFS bound ($\nu\!>\!2/d$) for random systems, which suggests that the MBL transition in 1D QP systems is Harris stable against weak quenched randomness. It would be interesting to study MBL transitions and their universal properties in higher-dimensional QP systems by employing this RSRG approach in the future.

{\it Acknowledgement}:  We would like to thank Shao-Kai Jian, Zi-Xiang Li and Yi-Fan Jiang for helpful discussions. This work is supported in part by the MOST of China under Grant No.
2016YFA0301001 (S.-X.Z. and H.Y.) and the NSFC under Grant No. 11474175 (S.-X.Z. and H.Y.). We also acknowledge the Guangzhou Tianhe-II supercomputer center for computational support.

\begin{widetext}
	
\section{Supplementary Material}
\renewcommand{\theequation}{S\arabic{equation}}
\setcounter{equation}{0}
\renewcommand{\thefigure}{S\arabic{figure}}
\setcounter{figure}{0}

\subsection{A. Details of the improved RSRG approach}
We now describe in details the implementation of the improved real-space renormalization group (RSRG) approach in 1D QP systems and discuss the effectiveness of this method in studying critical behaviors of MBL transitions. The main philosophy of such analysis is to study the structure of so-called resonance clusters. A resonance cluster can be understood as a region consisting a number of sites where a wave packet can tunnel back and forth hence contributes destabilization to the overall MBL nature. The MBL phase can be destroyed as long as there are large enough resonance clusters. However, identifying all such generic many-body resonances rigorously is as difficult as exactly solving the corresponding many-body problem, i.e. which is as limited to small size as in exact diagonalization studies. But around the MBL transition point, it is reasonable to assume that the structure of critical resonant clusters exhibit certain self-scaling geometry at different scales due to the criticality. Such hierarchical structure allows a much simpler way of identifying resonances instead of solving the full many-body model by exact diagonalization.

Identifying resonances plays key roles in understanding MBL delocalization transition and underlines main principles of real space RG. Instead of solving the full resonance structure from the beginning, we first identify small resonant clusters (starting from two-sites resonance pairs), and then examine whether groups of these small resonant clusters can collectively resonate on larger space or longer time scale. By applying this analysis iteratively, we could finally reveal the structure of resonance clusters in the large scale. Moreover, the above procedures cost much less computation time than directly solving the problem by exact diagonalization from beginning, which is practically impossible for any size larger than $\sim 20$ sites.

To summarize, we intend to capture the universal property of MBL criticality by avoiding the full quantum microscopic problems and instead ask help from some kind of RG process. In other words, the problem is reduced to answer several detailed questions of such real-space RG method: 1) Which parameter to track in the iterative process; 2) How to start the process (i.e. the initial value of characteristic parameters); 3) How about the iterative steps (i.e. rules of grouping clusters into larger ones based on certain parameters and also rules of renormalization of parameters themselves).

To illustrate the approach in a more concrete way, we consider the following Hamiltonian given by
\eq{H=H_0(t,\mu)+H_{int}(V),}{arg2}
where $H_0$ is the non-interacting part with random variables $\mu$ and $H_{int}$ is the interaction part of the Hamiltonian. To determine whether resonance occurs, we need to compare tunnel matrix elements between different sites which feature thermal nature with the energy mismatch between different sites which favors localization nature. We denote tunneling matrix elements as $\Gamma_{ij}$, representing the typical tunneling amplitude between $i$ and $j$ sites (or clusters). As interactions between neighboring sites contribute to thermal nature and enhance the tunneling, we have $\Gamma_{ij}=V e^{-r_{ij}/\xi}$ as the initial values, where $\xi$ is the localization length and $r_{ij}$ represents the {\it effective} distance between $i$ and $j$ sites (clusters). We use the single-particle localization length as an approximation in the calculation. Note that $r_{ij}$ may be different from the real distance between sites $i$ and $j$, since $\Gamma$ is the matrix element between eigenstates instead of sites. We will elaborate on this subtlety more later as this is very important being one of our major improvements to the original RSRG proposal when studying MBL transitions in QP systems. For the characteristic parameter of energy mismatch, we denote it as $\Delta E_{ij}$ representing the typical energy mismatch between $i$ and $j$ sites(or clusters). In the original proposal of RSRG, the difference in random chemical potentials between $i$ and $j$ sites is chosen as the initial value of energy mismatch. This choice looks reasonable but does not work well for the QP systems. In our improved algorithm, we choose a different scheme for setting this initial value which we shall discuss in details below.

We now discuss in details how we improve the RSRG scheme in studying MBL transitions of QP systems. The improvements mainly consist of choosing initial values in different ways. By such improved scheme we can take more microscopic details of the interacting model into consideration. For each disorder configuration (namely sampling the random variables based on given probability distribution in $H_0$), we first carry out exact diagonalization calculation of the non-interacting part. We consider $H_0$ which features Anderson localization. Then, we can obtain the single-particle energy spectrum (eigenvalues) and corresponding wave functions (eigenstates) for this non-interacting part. For the case of full Anderson localization, all single-particle wave functions are localized in real space and we can calculate the localization center of each single-particle wave function and use it as the characteristic position of the wave function: $r(i)$. Instead of using the chemical potential and the real position of sites in the original RSRG proposal, we employ the characteristic positions $r(i)$ of single-particle wave functions and corresponding real energy eigenvalue $E(i)$ as the initial input of our improved RSRG algorithm. This improvement makes the algorithm more universal and accurate. Especially, such preprocessing is necessary for system with QP potentials. Though the original simple approximation works well in systems with random potentials it fails for the case of QP potential. The reason for the difference is based on the distinct behaviors of wave functions and energy spectrum between the two cases in the presence of the hopping terms, as we articulate below.

When the hopping $t=0$, the non-interacting system is fully localized; onsite potential and site position can be used as accurate input data for RSRG process. When hopping amplitudes are finite ($t\neq 0$), the single-particle wave-functions now have certain broadening across many sites. Moreover, energy spectrums of wave-functions are shifted from the value of onsite potentials. For the case of quenched random disorder, such shifts can be viewed as adiabatic change. However, such approximation fails for the case of QP potential. The spectrum for single-particle energy cannot be adiabatically connected to onsite potential due to the gap in the spectrum of the AA model, which is in stark contrast with the case of quenched random potential which has the continuous spectrum even in the presence of finite hopping amplitudes. In other words, when $t=0$, the energy spectrum of the non-interacting part of the Hamiltonian are simply the onsite chemical potentials, which are gapless for both cases of uniform distributed random variable and QP potential. But, when the hopping $t\neq 0$,  the energy spectrum for the case of random potential is still gapless; this fact allows the above explanation as an adiabatic approximation. Adiabatic approximation justifies the success of the original RSRG approach for the random potential. However, the energy spectrum for the case of QP potential has a finite gap when $t\neq 0$. We think that this should ruin applicability of the adiabatic approximation and lead to a failure of the original RSRG algorithm when applied to the QP system. Indeed, the failure was encountered when we applied the original RSRG algorithm to study the QP problem.

After we obtain information about the (characteristic) position and energy level of each single-particle wave function, we can use the initial values of these two key quantities to start the RG iterative process: tunneling strength $\Gamma$ and energy-level mismatch $\Delta E$. The RSRG iterative process depends on two characteristic parameters between every two clusters: tunneling amplitude $\Gamma_{ij}$ (namely matrix element for transitions between different clusters) and typical energy mismatch $\Delta E_{ij}$ (namely many-body level spacing for the new cluster consisting of clusters $i$ and $j$). The key iterative procedure of such RG is to merge all clusters with $\Gamma_{ij}\!>\!\Delta E_{ij}$ in each step, and then calculate new sets of renormalized parameters $\Gamma'$ and $\Delta E'$ by corresponding formulas for newly generated cluster structures. Specifically, their initial values are given as follows: $\Gamma^0_{ij}=V \exp[-\frac{\abs{r(i)-r(j)}}{\xi}]$ and $\Delta E^0_{ij}=\abs{E(i)-E(j)}$, where $\xi$ is the single-particle localization length and $V$ is the characteristic value of interactions in $H_{int}$. For the case of the AA model with QP potential, the localization length is given by $\xi = \frac{1}{\log W/2}$, where $W$ represents the strength of the QP potential.

The next step is to run the iterative process of RSRG. Initially, the localization center of each single-particle wave function is recognized as the position of initial clusters. By comparing the tunneling strength and energy mismatch between clusters $i$ and $j$, a resonating bond between $i$ and $j$ is assigned if $\Gamma_{ij}>\Delta E_{ij}$. During each RG step, all clusters connected by a path of resonating bonds are merged into a new larger cluster. Therefore, new data of $\Gamma, \Delta E$ can be derived based on the iterative formula. For new energy mismatch value, we have
\al{\Lambda_{i'}&=\sqrt{\sum_i\Lambda_i^2+\sum_{ij}\Gamma_{ij}^2}\\
	\delta_{i'}& = \Lambda_{i'}/(2^{n_{i'}}-1)\\
	\Delta E_{i'j'}&=\delta_{i'}\delta_{j'}/\min(\Lambda_{i'},\Lambda_{j'}),}
where $i',j'$ label newly formed clusters, $n_{i'}$ is the number of sites contained in the corresponding cluster $i'$,  and $\Lambda$ is the assistance parameter we track whose role are similar to effective chemical potential. To obtain  $\Lambda_{i'}$ for each newly formed cluster $i'$, we sum over inner $\Lambda_i$ and off-diagonal matrix element $\Gamma_{ij}$, where $i,j$ are old clusters constituting the new cluster $i'$. Note that there is an exception for the last formula; when $\Lambda_{i'}\geq \delta_{i'}\geq\Lambda_{j'}\geq\delta_{j'}$, we take $\Delta E_{i'j'}=\max(\delta_{i'}-\Lambda_{j'},\delta_{j'})$ instead. All these formula for $\Delta E$ of newly generate clusters can be justified straightforwardly by considering merge two systems with energy levels together and determine the new typical energy gap.

We now discuss the the renormalization of $\Gamma$ during the iterative process. If two clusters are not modified during a RG step, the coupling between them is set to zero, which reflects the nature of MBL. If at least one of the two clusters is modified during the RG step, $\Gamma$ is given by
\eq{\Gamma_{i'j'}=\max(\Gamma_{ij})e^{-sth/2(n_{i'}+n_{j'}-n_i-n_j)},}{arg2}
where $sth = \ln 2$ is the characteristic entropy per site in the thermal phase.
Namely, we choose the largest $\Gamma$ from the original clusters forming the two new clusters and normalize it as the new $\Gamma$. This form is believed to approximate the resonating clusters as small locally thermal sub-systems which reflects ETH principles. See Ref. 45 
of the main text for more details on the iterative formula; we adopt their conventions and implementation on the iterative process of RSRG.

The RG iterative process terminates when no new resonant bond forms, i.e. reaching the fixed point where the cluster structure receives no modifications by further RG steps. Now, one obtains the fixed point cluster structure for each disorder configuration. We can carry out measurements on the normalized entanglement entropy $s$ in this system based on the final cluster structure obtained after RG. The normalized EE is calculated by $s=\frac{\sum_{C}\min(C_L,C_R)}{L/2}$, where we sum over all final clusters $C$ crossing the half cut of the system and $C_{L/R}$ represents the number of sites in the left/right half of the cluster $C$. This formula is based on the fact that thermal phases feature volume law for entanglement entropy. After we calculate the normalized entanglement entropy $s$ for each disorder configuration, we can perform the disorder average to obtain $\bar{s}$.

We performed RSRG calculations for different Hamiltonian parameters and different system size $L$. By collecting all data, we can apply finite-size scaling analysis to extract critical exponents. The guiding scaling formula is $s=f[(W-W_c)L^{1/\nu}]$, where the critical potential $W_c$ can be located by the crossing point of half-chain entanglement entropy density obtained for different system sizes, and the critical exponent $\nu$ can be extracted by standard scaling collapse. For a more systematic and quantitative approach to extract critical exponent, see the finite-size scaling analysis method in the appendix of Ref. 64 
of the main text. We adopt their strict analysis on the finite-size scaling by taking the errorbar of entanglement density from our RG calculation into consideration, which gives much more precise estimation on the critical exponents compared to simple finite-size data collapse.

\subsection{B. More results on stability of QP-MBL criticality}

To further confirm that the MBL transition induced by quasi-periodic potential is stable against weak quenched randomness, we applied the improved RSRG calculations to QP systems in the presence of on-site randomness potential ($W'=0.1t$). In the presence of the weak quenched randomness, our calculations give rise to the critical QP strength $W_c=3.32\pm0.02$ and the critical exponent $\nu = 2.5\pm 0.3$. The results are shown in \Fig{qprandom}. The result of critical exponents in the presence of weak quenched randomness matches quite well with the one obtained without the weak quenched randomness, which implies that weak quenched randomness does not change the universality class of MBL transition induced by pure QP potential in the absence of quenched randomness. This fact directly confirms that such QP-MBL criticality is actually stable against weak quenched randomness, namely the weak randomness is Harris-irrelevant.

\ipic{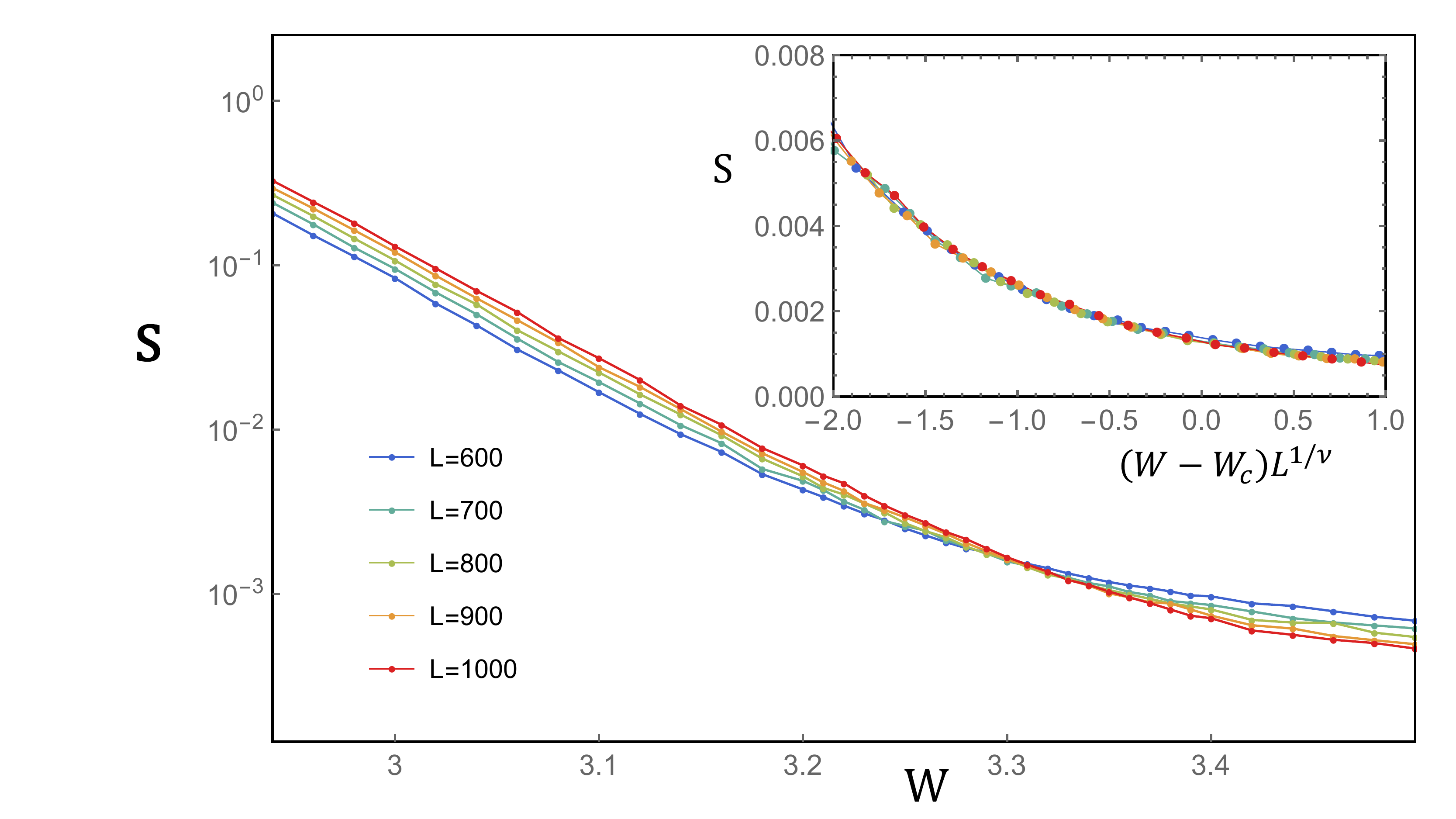}{Finite-size scaling analysis of entanglement entropy density of 1D systems with quasiperiodic potentials and small extra quenched disorder. $W_i\!=\!W\cos(2\pi \alpha i+\phi)$ and on-site randomness $W'=0.1t$ at the same time. The MBL transition is identified as the crossing point of entanglement entropy $s$ for different sizes $W_c\!=\!3.32\pm0.02$. The data collapse shown in the inset gives rise to $\nu\!=\!2.5\pm0.3$. }{qprandom}{width=7cm}

\subsection{C. Discussions on standard deviation of entanglement entropy}
The standard deviations of the half-chain entanglement entropy (EE) can be evaluated by jack-knife resampling method from the RSRG approach. We studied the size dependence of the standard deviations of the half-chain EE across samples. For both cases of QP potential and quenched randomness, our calculations show that the size dependence of the standard deviation of the half-chain EE is quite weak for the size $L$ in the range of $600$$\sim$$1000$, indicating that our calculations indeed access critical behaviors of the MBL transition. For both random and QP cases, the standard deviations are nearly $0.5$ around the critical value of disorder strength $W$. Although the $W$-dependence of the standard deviations in our RSRG studies is similar to the one obtained with smaller size in Ref. 42, the peak value of the standard deviations in our studies is higher than the one obtained from the exact diagonalization (ED) studies in Ref. 42. For the QP case, from our RSRG calculations we found that the standard deviations for size $L=600$$\sim$$1000$ are much higher than the value obtained by ED for size $L=12$$\sim$$18$. This indicates that the standard deviations obtained in ED calculations for small size are not yet sufficiently close to the critical scaling regime. If the ED calculation were able to compute the standard deviation for much larger system size $L$, a larger value of relative deviation close to $0.5$ might be obtained. In other words, it means that for the QP case the system size $L=12$$\sim$$18$ studied in ED calculations might be somewhat far from sufficiently good scaling regime. Therefore, it is reasonable to believe that the results obtained from our RSRG calculations of size about $1000$ sites should be sufficiently persuasive and reliable in deriving critical behaviors of the MBL transition.

\end{widetext}
\end{document}